\documentclass[sigconf,nonacm,natbib=false]{acmart}

\usepackage{listings}
\usepackage{physics}
\usepackage{algorithm}
\usepackage{algpseudocode}
\usepackage{subcaption}

\definecolor{codegreen}{rgb}{0,0.6,0}
\definecolor{codegray}{rgb}{0.5,0.5,0.5}
\definecolor{codepurple}{rgb}{0.58,0,0.82}
\definecolor{backcolour}{rgb}{0.95,0.95,0.92}

\lstdefinestyle{mystyle}{
    backgroundcolor=\color{backcolour},   
    commentstyle=\color{codegreen},
    keywordstyle=\color{magenta},
    numberstyle=\tiny\color{codegray},
    stringstyle=\color{codepurple},
    basicstyle=\ttfamily\footnotesize,
    breakatwhitespace=false,         
    breaklines=true,                 
    captionpos=b,                    
    keepspaces=true,                 
    numbers=left,                    
    numbersep=5pt,                  
    showspaces=false,                
    showstringspaces=false,
    showtabs=false,                  
    tabsize=2,
    language=c++ 
}

\usepackage{xcolor} 
\newcommand{\Redacted}[1]{\textbf{redacted for double blind}}
\renewcommand{\Redacted}[1]{#1}

\usepackage{algorithm}
\usepackage{algpseudocode}

\RequirePackage[
  datamodel=acmdatamodel,
  style=acmnumeric,
  ]{biblatex}

\begin{document}

\title{GPU Acceleration of Monte Carlo Tallies on Unstructured Meshes in OpenMC with PUMI-Tally}

\author{Fuad Hasan}
\email{hasanm4@rpi.edu}
\affiliation{%
  \institution{Rensselaer Polytechnic Institute}
  \city{Troy}
  \state{New York}
  \country{USA}
}

\author{Cameron W. Smith}
\email{smithc11@rpi.edu}
\affiliation{%
  \institution{Rensselaer Polytechnic Institute}
  \city{Troy}
  \state{New York}
  \country{USA}
}

\author{Mark S. Shephard}
\email{shephard@rpi.edu}
\affiliation{%
  \institution{Rensselaer Polytechnic Institute}
  \city{Troy}
  \state{New York}
  \country{USA}
}

\author{R. Michael Churchill}
\email{rchurchi@pppl.gov}

\affiliation{%
  \institution{Princeton Plasma Physics Laboratory}
  \city{Princeton}
  \state{New Jersey}
  \country{USA}}
  
\author{George J. Wilkie}
\email{gwilkie@pppl.gov}

\affiliation{%
  \institution{Princeton Plasma Physics Laboratory}
  \city{Princeton}
  \state{New Jersey}
  \country{USA}}

\author{Paul K. Romano}
\email{promano@anl.gov}
\affiliation{%
  \institution{Argonne National Laboratory}
  \city{Lemont}
  \state{Illinois}
  \country{USA}
}

\author{Patrick C. Shriwise}
\email{pshriwise@anl.gov}
\affiliation{%
  \institution{Argonne National Laboratory}
  \city{Lemont}
  \state{Illinois}
  \country{USA}
}

\author{Jacob S. Merson}
\email{mersoj2@rpi.edu}
\orcid{0000-0002-6813-6532}
\authornote{Corresponding Author}
\affiliation{%
  \institution{Rensselaer Polytechnic Institute}
  \city{Troy}
  \state{New York}
  \country{USA}
}
%
%
%
%
%
%
%

\renewcommand{\shortauthors}{\Redacted{Hasan} et al.}


\begin{abstract}
Unstructured mesh tallies are a bottleneck in Monte Carlo neutral particle transport simulations of fusion reactors. This paper introduces the PUMI-Tally library that takes advantage of mesh adjacency information to accelerate these tallies on CPUs and GPUs. For a fixed source simulation using track-length tallies, we achieved a speed-up of 19.7X on an NVIDIA A100, and 9.2X using OpenMP on 128 threads of two AMD EPYC 7763 CPUs on NERSC Perlmutter. On the Empire AI alpha system, we achieved a speed-up of 20X using an NVIDIA H100 and 96 threads of an Intel Xenon 8568Y+. Our method showed better scaling with number of particles and number of elements. Additionally, we observed a 199X reduction in the number of allocations during initialization and the first three iterations, with a similar overall memory consumption. And, our hybrid CPU/GPU method demonstrated a 6.69X improvement in the energy consumption over the current approach.
\end{abstract}




\maketitle

\section{Introduction}

Monte Carlo particle transport methods will play a particularly important role in the enablement of grid-connected fusion power plants. In particular, neutronics calculations are needed for radiation shielding and blanket design and neutral particle calculations are needed to understand neutral recycling and neutral beam injection.

Monte Carlo simulation tools have coalesced around the use of constructive solid geometry (CSG) modeling techniques. This is an effective strategy because it supports fast point localization. However in fusion reactors, the geometry is considerably more complex than in fission reactors requiring a rethinking of traditional tooling. Two options currently exist, to hand-code CSG models, such as with the heroic effort undertaken to construct neutronics models for ITER \cite{juarezFullHeterogeneousModel2021,leichtleITERTokamakNeutronics2018}, or perform mesh-based Monte Carlo simulations. Mesh-based methods can use off the shelf automated tools to rapidly integrate updates in underlying CAD models. Unfortunately, unstructured mesh-based particle method calculations are historically slow since they require many particles for adequate statistics in each element, and because the current design of Monte Carlo codes requires re-localization in each element.

Over the past few years, substantial effort has gone into the development of strategies to effectively take advantage of CAD geometries with Monte Carlo neutronics simulations. These efforts have revolved around a few strategies: to construct surface meshes from step files \cite{CAD_to_OpenMC2025,shimwellCADDAGMCConvert2024, Stellarmesh2023} and perform ray tracing with a tool such as DAGMC \cite{wilsonAccelerationTechniquesDirect2010}, to construct a volume mesh and trace particles through the mesh \cite{burkeVerificationUnstructuredMesh2012,martzNotableComparisonComputational2013,zavorkaUnstructuredMeshBased2023,zhangVolumeMeshBased2025}, or try to reverse engineer a CSG tree from a BREP, a task that has no unique solution \cite{duInverseCSGAutomaticConversion2018, harbNovelAlgorithmCAD2023,luoMeshCNNbasedBREPCSG2022}.

To our knowledge, this work represents the first GPU accelerated volume tally system for neutral particle Monte Carlo codes. However, our results emphasize that our structured approach to geometry handling and effective use of cache coherent data structures present significant performance advantages over the state-of-the art for handling of unstructured mesh tallies.

Unstructured mesh calculations offer an additional benefit over CSG models: they more closely match the design of engineering analysis codes which makes coupling easier \cite{novakMonteCarloMultiphysics2024}. Mesh based coupling can be accomplished through interfaces such as the Parallel Coupler for Multimodel Simulation \cite{shephardUnstructuredMeshTools2024} or DTK \cite{slatteryDataTransferKit2013}.

In this paper, we introduce a library, PUMI-Tally\footnote{Available at \Redacted{https://github.com/Fuad-HH/PumiUMTally/}}, for performing GPU accelerated tally operations and its integration into the open source Monte Carlo code OpenMC \cite{romanoOpenMCStateoftheartMonte2015}. OpenMC is a community developed Monte Carlo code for neutron and photon transport simulations and has been targeted as a platform for the next generation of atomic transport and plasma interaction simulations to replace DEGAS2 \cite{stotlerNeutralGasTransport1994, wilkieDemonstrationOpenMCFramework2024}. PUMI-Tally takes advantage of adjacency based ray tracing and specialized particle data structures to accelerate Monte Carlo tally operations. It is based off prior developments of the PUMIPic infrastructure for hybrid particle mesh calculations \cite{diamondPUMIPicMeshbasedApproach2021}.

Contributions in this work include:
\begin{itemize}
    \item Development of a high-performance API for GPU accelerated neutronics tally operations based on PUMIPic.
    \item Demonstration of a heterogeneous application that utilizes both CPUs and GPUs to hide part of the simulation costs.
    \item A generalization of the PUMIPic search interface to separate construction of physics kernels from fundamental particle search and localization operations.
\end{itemize}


\section{Background}
Neutral particles are unaffected by electromagnetic fields and their transport is governed by the Boltzmann equation, an integro-differential equation which describes the mesoscale physics of the evolution of a distribution function.
The Boltzmann equation is given by
\begin{equation}
    \pdv{f}{t} + \vb{v} \cdot \nabla_{\vb{r}} f+ \vb{F} \cdot \nabla_{\vb{v}} f + S = \left( \pdv{f}{t} \right)_{\text{coll}},
    \label{eq:boltzmann}
\end{equation}
where \(f\) is the particle distribution function, \(\vb{v}\) is the particle velocity, \(\vb{r}\) is the particle position, \(F\) is the external force per unit mass, \(S\) is the particle source, and the right hand side represents the impact of collisions (scattering, and absorption). 


The difference in the solution methods for neutral particles and neutronics primarily revolve around the choice for the collision operator which specifies how two or more particles may interact as a function of particle position and velocity (or energy). Recent advances have demonstrated the implementation of neutral physics into OpenMC for hydrogen undergoing electron-impact ionization/excitation and charge exchange with protons \cite{wilkieDemonstrationOpenMCFramework2024}.

The collision term is an integral term over the six dimensional phase space. And, relates an incoming particle direction and velocity to an outgoing particle direction and velocity which is governed by the collisional cross-section. In the continuous energy neutronics setting, computing the cross-sections is complex and relies on experimental data. This complexity motivates our choices as to which parts of the algorithm continue to be performed by OpenMC and which parts are performed by PUMI-Tally.

Equation \eqref{eq:boltzmann} needs to be solved to find the distribution function at every location in phase space. This is made difficult by the high dimensionality of phase space (6D). Due to the high dimensionality, Monte Carlo methods have proven to be an effective method. However, due to the extraordinary cost of resolving the distribution function on a point-wise basis, it is resolved in small volumes in phase space. The phase space volumes are selected by a filter and the process of recording quantities into the filtered domain is called tallying.

From a practical perspective, there are two common methods to estimate the tallies: collision based, or track length based. Collision estimators are fairly straightforward, you record the filtered bin where a particle undergoes a collision. Track length estimators are more accurate with fewer particles, however they are more difficult to implement. The computation of a track length estimator requires measuring the length that a particle traversed each filter bin in phase space then summing that value, appropriately weighted, into the appropriate filter bin. When performing tallying on an unstructured mesh this requires computing the line integral of the particle trajectory across each mesh element.

\begin{algorithm}
    \caption{Event-Based Monte Carlo Method}\label{alg:event-based}
    \begin{algorithmic}
        \For{batch in nbatches}
        \While{batch not done}
            \State Initialize event queues with particles
            \While{Any event queue has particles}
                \State Execute event for biggest queue
            \EndWhile
        \EndWhile
    \EndFor
    \end{algorithmic}
\end{algorithm}

\begin{algorithm}
    \caption{Advance Particle Event}\label{alg:advance}
    \begin{algorithmic}
        \For{All particles in the "advance particle" queue}
            \State $l_g \gets$ distance to geometric interface along particle trajectory
            \State $l_c\gets$ trial collision distance
            \State Particle destination $\gets \min(l_g, l_c) \hat{\vb{r}}$
            \State Score the tracklength tallies along the particle path
        \EndFor
    \end{algorithmic}
\end{algorithm}

\section{Implementation}

OpenMC supports two modes of transport: first, history-based, where each particle is tracked for its entire life while it undergoes events such as collisions, absorption, leakage, reflection, etc. In this mode, behavior of each particle is completely independent which leads to a trivially parallelizable set of algorithms that cannot be easily vectorized. In the second mode, called event-based, particles are grouped and each group of particles undergoes one type of event at a time. This method leads to vectorizable algorithms that support GPU computations and what is used in this work and shown in algorithm~\ref{alg:event-based} \cite{bleileInvestigationPortableEventBased,brownMonteCarloMethods1984,trammPortableGPUAcceleration2022}.

This work builds off of PUMIPic, which has been developed to support distributed calculations that include both particles and meshes \cite{diamondPUMIPicMeshbasedApproach2021,shephardUnstructuredMeshTools2024}. And it has been the basis for two plasma simulation codes and a code to model hypersonic flows \cite{nath3DUnstructuredMesh2023,nathGPUAccelerated3DUnstructured, zhangComet3DGPUAcceleratedThreeDimensional, zhangDevelopmentUnstructuredMesh2023}. PUMIPic takes a mesh-centric approach where particles maintain classification to owning mesh elements. PUMIPic makes use of the Omega\_h mesh data structure which maintains mesh adjacencies and classification to underlying geometric models and has been designed to be effective on GPUs \cite{ibanezCONFORMALMESHADAPTATION}. Both PUMIPic and Omega\_h utilize the Kokkos performance portability library to support running on the range of CPU and GPU systems \cite{carteredwardsKokkosEnablingManycore2014}.

In neutral particle transport, the geometric classification data can be used to determine when particles pass from one material region to another and can help to avoid resampling particle trajectories at every element intersection as is done in many mesh-based transport codes. In the context of tallies, the classification information can be used to aggregate values over regions of interest such as divertor tiles.

PUMIPic has been designed to support a common set of operations for simulations with both particles and meshes, however different data structures are needed depending on the number of particles in each element, the distance traveled by each particle in a step, and if particle collision operators require information about neighbor particles. In the case with particle-particle interactions, high numbers of particles, and a low mean free paths it is desirable to take advantage of cache locality by storing particles locally in each element. Three particle data structures described in \cite{diamondPUMIPicMeshbasedApproach2021}, SCS, CSR, and CabM handle this case being developed on top of Kokkos Views and the Cabana AoSoA data structures respectively \cite{carteredwardsKokkosEnablingManycore2014,Slattery2022}.

However, neutral particle simulations tend not to fall into this category with long mean free paths (compared to element sizes), low numbers of particles per element, and linear collision operators. In this case, we make use of the Disorganized Particle Data-structure (DPS) which maintains the particle to mesh mapping through a single integer value \cite{shephardUnstructuredMeshTools2024}. This limits the cost of rebuilding the particle data structures, but at the cost of reduced efficiency and cache coalescing when accessing element information that's needed for particle path tracing.

\begin{figure}
    \centering
    \includegraphics{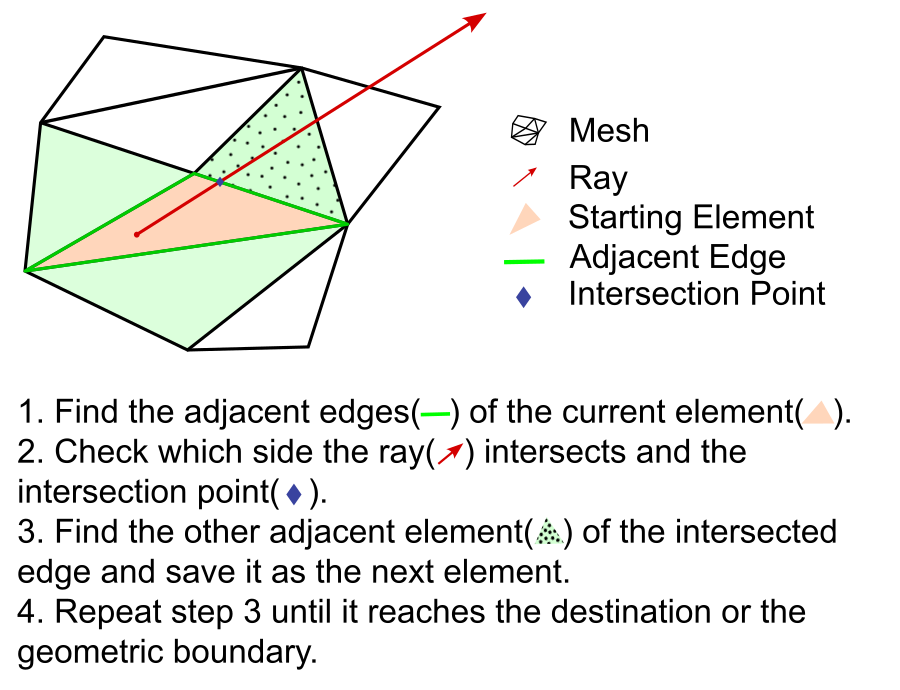}
    \caption{Adjacency based search algorithm.}
    \label{fig:adj_search_alg}
\end{figure}

Particle path tracing is done through an adjacency search method described in figure~\ref{fig:adj_search_alg} and Algorithm~\ref{alg:adj-search}. The particle destination is governed by complex physics and is performed on the CPU by OpenMC's current methods. Once the particle destination is chosen, it is copied from the OpenMC particle data structures into a host array. This host array is copied to a device array by PUMI-Tally and loaded into the PUMI particle data structures. At this point, the adjacency search proceeds. Finding the ray-face intersection points in simplex meshes can be made quite efficient by making use of the barycentric coordinates. If the destination is not within the current element, we loop through the three element faces that the particle did not enter from and check if the ray face intersection returns a valid parametric coordinate along the ray. For straight sided 3D elements, this corresponds to solving a $3 \times 3$ system of equations (one barycentric coordinate can be eliminated since we know we are on an element face). Once the particle knows which face it will exit through, adjacency information can be used to retrieve the next element the particle will enter. All particles simultaneously move across one element at a time.

Before the particle is updated to the next element, a user callback is called by the search routine. This callback, shown in listing~\ref{lst:intersection_functor}, is quite general and allows the user to control the particle behavior, including setting a different destination than the one expected by the adjacency search and killing particles or resurrecting them if they were killed previously. The search is templated on the callback type, so the user can pass in a functor that stores additional information that's needed for performing physics operations or tallies.

In this work, the functor stores GPU arrays that are sized by the number of elements times the number of energy bins that the user specifies for tallies. The size of these arrays is known during program initialization and can be allocated up front. In this way we perform tallies as the particles are transported across the mesh as tallying simply requires storing a cumulative sum into the appropriate bin as shown in equation \eqref{eq:tally}.

\begin{equation}
    X=\iint g(\vb{r}, \vb{v}) f(\vb{r}, \vb{v})\dd{ \vb{r}} \dd{\vb{v}},
    \label{eq:tally}
\end{equation}
where, \(g\) is some quantity (scoring function in OpenMC) and \(f\) is the particle distribution function.

Note that unlike other methods, we do not require expensive particle-in-element localization procedures that are often implemented as trees that perform poorly on the GPU. One place where localization is required is during particle initialization. We expected to make use of a uniform grid search method that has been implemented for Omega\_h meshes on the GPU in the Parallel Coupler for Multimodel simulations \cite{shephardUnstructuredMeshTools2024}. However, initial testing was done by placing all particles at a trial initial location that was chosen as one of the element centroids, then using the adjacency search to move the particle to its true starting point. We found that this method was satisfactory as it did not represent a large percentage of our runtime.

\begin{lstlisting}[caption=Element interface callback API, label=lst:intersection_functor]
struct IntersectionFunctor {
operator()(const Mesh & mesh,
            ParticleStructure<ParticleType> * particles, // includes position and destination
            Read<LO> current_element_ids &,
            Read<LO> & exit_face, // face where particle will exit element
            Read<Real> & intersection_points, // point where the particle intersected the face, or the destination
            Write<LO> & next_element_ids, // id of the element the particle should move into
            Write<bool> & particle_done, // if the particle should continue moving, or be resurected
            );
};
\end{lstlisting}

\begin{algorithm}
    \caption{Adjacency Based Search Algorithm}\label{alg:adj-search}
    \begin{algorithmic}
        \While{The particle does not reach its destination}
            \State Get the adjacent sides of its current element
            \For{All sides of the current element}
                \State{Check if the particle path intersects the side}
                \If{Intersects}
                    \State Save the side and find the intersection point
                \EndIf
            \EndFor
            \State Get the adjacent elements of the intersected sides
            \If{The side has 2 adjacent elements}
                \State Move the particle to the other element
            \ElsIf{The side has 1 adjacent element}
                \State The particle has reached the boundary
                \State Apply boundary condition
            \EndIf
        \EndWhile
    \end{algorithmic}
\end{algorithm}

\section{Integration into OpenMC}
Our initial approach to integrate unstructured mesh tallies with OpenMC was to directly call PUMIPic based functionality within OpenMC. However, this led to significant difficulties when building with GPU compilers and would have required a significant rewrite of portions of OpenMC to be able to compile with the Kokkos provided \texttt{nvcc\_wrapper}. To address this challenge, we separated the tally operations into the independent PUMI-Tally library. PUMI-Tally utilizes the pointer-to-implementation (PIMPL) idiom which affords the construction of a \texttt{PumiTally} object without the need to link to any of the dependency libraries (PUMIPic, Kokkos, etc.) to OpenMC, or build OpenMC with a GPU compiler. The PIMPL idiom does require a pointer indirection, however PUMI-Tally utilizes batched interfaces (listing~\ref{lst:pumi-tally-interface}), so the cost of the pointer indirection is small compared to the amount of work that is performed for each call.

The interface for PUMI-Tally, outlined in listing~\ref{lst:pumi-tally-interface} is relatively simple requiring a constructor that loads a Omega\_h mesh, and initializes the PUMIPic data structures with a set number of particles. A function to initialize particles that is called at the start of a batch to perform localization of a particle to its initial parent element. And, a function to update a particles destination, weights, and if it should be moving during the current iteration. Lastly, a function to write the tallies to disk in Omega\_h, VTK, or Adios2 file formats.

To enable the data parallel tally approach used here, OpenMC is run in the event based mode summarized in algorithm~\ref{alg:event-based}. This approach, originally developed to support vector computers \cite{brownMonteCarloMethods1984}, groups particles into event queues where each particle in a queue will undergo the same type of operation at once. This is quite different than the more typical history based approach where each particle is transported independently for its entire lifetime. Events that are processed in OpenMC include calculating material cross-sections, advancing the particle locations, sampling new trajectories when particles cross geometric boundaries, and computing particle collisions.

The unstructured mesh tallies are computed during the advance particle location event described in algorithm \ref{alg:advance}. The first three steps of this algorithm refer to calculating the distance to the next geometric interface and sampling the distance to the next collision to obtain the final particle location. This part of the process is called transport and relies on complex material information. The last step of the advance particle location event is computing the tallies. In the current work, transport is performed by OpenMC on the CPU and the tallies are performed by PUMI-Tally either on the CPU or GPU depending on the selected backend.

Although the particles are batch processed in the event based method, particle location and weight data is stored in a struct that defines the data for each particle. To utilize PUMI-Tally, the data must first be copied from the OpenMC particle data structures to a host array. This array is passed to the PUMI-Tally library where it is copied from the host to the device if utilizing a GPU backend. And, the data is copied from the device array to a PUMIPic particle data structure on the device. The storage mechanism of the PUMIPic particle data structure will depend on the storage method chosen (DPS, SCS, CSR, CabM). In the case of the DPS particle data structure used here, the particle information is stored in a Cabana AoSoA.

\begin{lstlisting}[caption=PUMI-Tally Public Interface, label=lst:pumi-tally-interface]
PumiTally(std::string &mesh_filename, int64_t num_particles, int &argc, char **&argv);
void initialize_particle_location(double *init_particle_positions, int64_t size);
void move_to_next_location(double *particle_destinations, int8_t *flying, double *weights, int64_t size);
void write(std::string& filename);
\end{lstlisting}

\section{Problem Setup and Verification}
\begin{figure}
    \centering
    \includegraphics[width=0.7\linewidth]{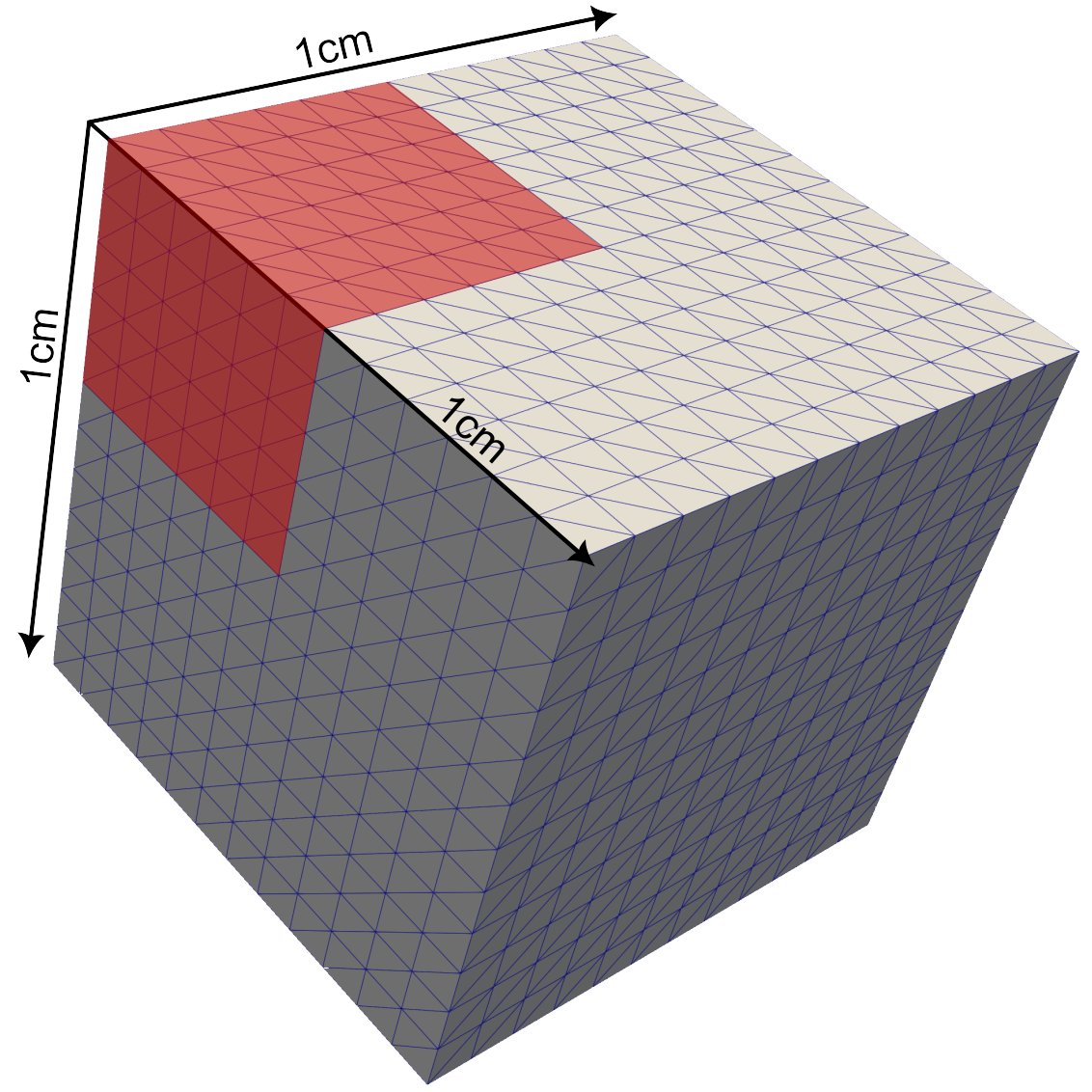}
    \caption{Simulation domain for verification problem. The red cube represents the location of the neutron source.}
    \label{fig:mesh}
\end{figure}
In this study, we perform a fixed source calculation in a cube domain with edge lengths of 1~cm. The source, is applied in 1/8th of the model shown in figure~\ref{fig:mesh}. Vacuum boundary conditions are applied to the outer surfaces of the cube. These boundary conditions mean that as particles hit the boundary they leave the simulation domain and are killed. For the verification problem, all neutrons are simulated with a multigroup cross section with a single energy. And, only scattering is performed. The macroscopic scattering and total cross section are the same given by ($\Sigma_t=\Sigma_s = 100\,\text{cm}^{-1}$). This corresponds to a particle mean free path of \(1/100\,\text{cm}\). Although we pick a simple cross-section here for verification, we note that since transport is being performed by OpenMC, we can perform tallies for continuous energy simulations with no modifications to the current approach.

\begin{figure}
    \centering
    \includegraphics[width=0.9\linewidth]{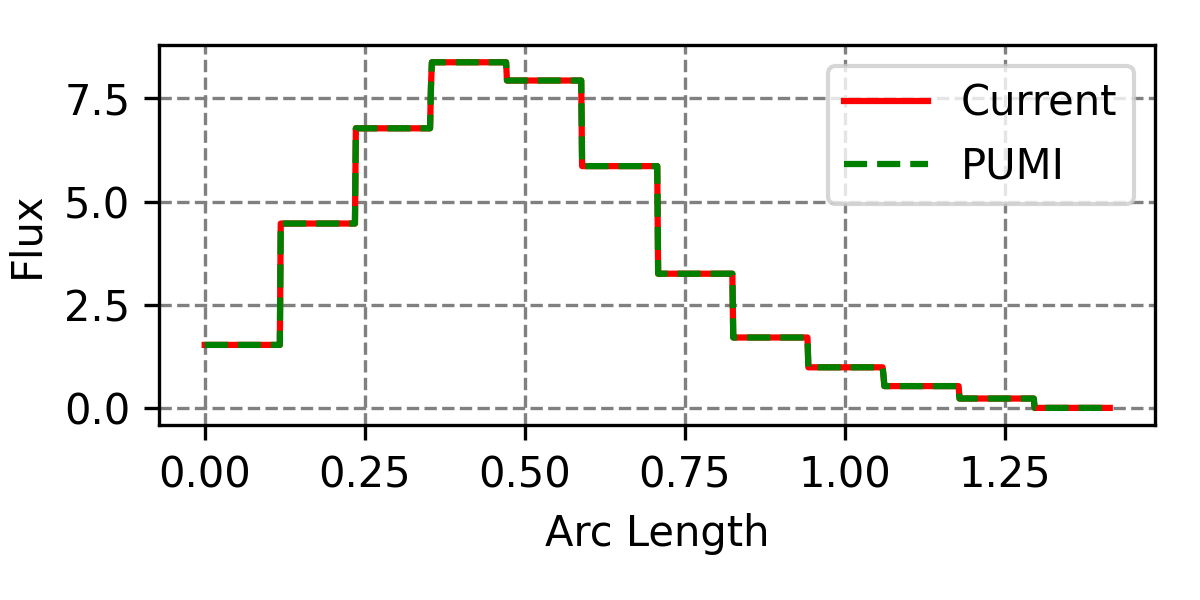}
    \caption{Comparison of flux tallies from current OpenMC and our version of OpenMC utilizing PUMI-Tally.}
    \label{fig:verification}
\end{figure}

To verify the solution, we compare the tallies recovered from the current version of OpenMC to our modified version of OpenMC that utilizes PUMI-Tally. Figure~\ref{fig:verification} shows the recovered flux along a line. The flux looks like a step function because the flux is piecewise constant within each element. We see that there is no difference between the recovered flux values.


\section{Results}
Performance measurements are done on two different computing platforms. NERSC Perlmutter and Empire AI alpha. Each Perlmutter node has 2 AMD EPYC 7763 CPUs with 128 total cores and 4 NVIDIA A100 with 40GB of memory. Code was compiled with the cray compilers and Cuda 12.4. Empire AI alpha has 8 NVIDIA H100 GPUs with 80GB of memory and 2 Intel Xenon 8568Y+ with 96 total cores. On Empire AI, GCC 13.1.0 with Cuda 12.4 was used.

To simplify the analysis, all presented results are computed using a single GPU. Since OpenMC utilizes replicated geometry on each rank, every process handles an independent set of particles that do not need to interact. Therefore, the tallies are trivially parallelizable onto multiple GPUs requiring a single all gather before writing to disk.

\begin{figure}
    \centering
    \includegraphics[width=1\linewidth]{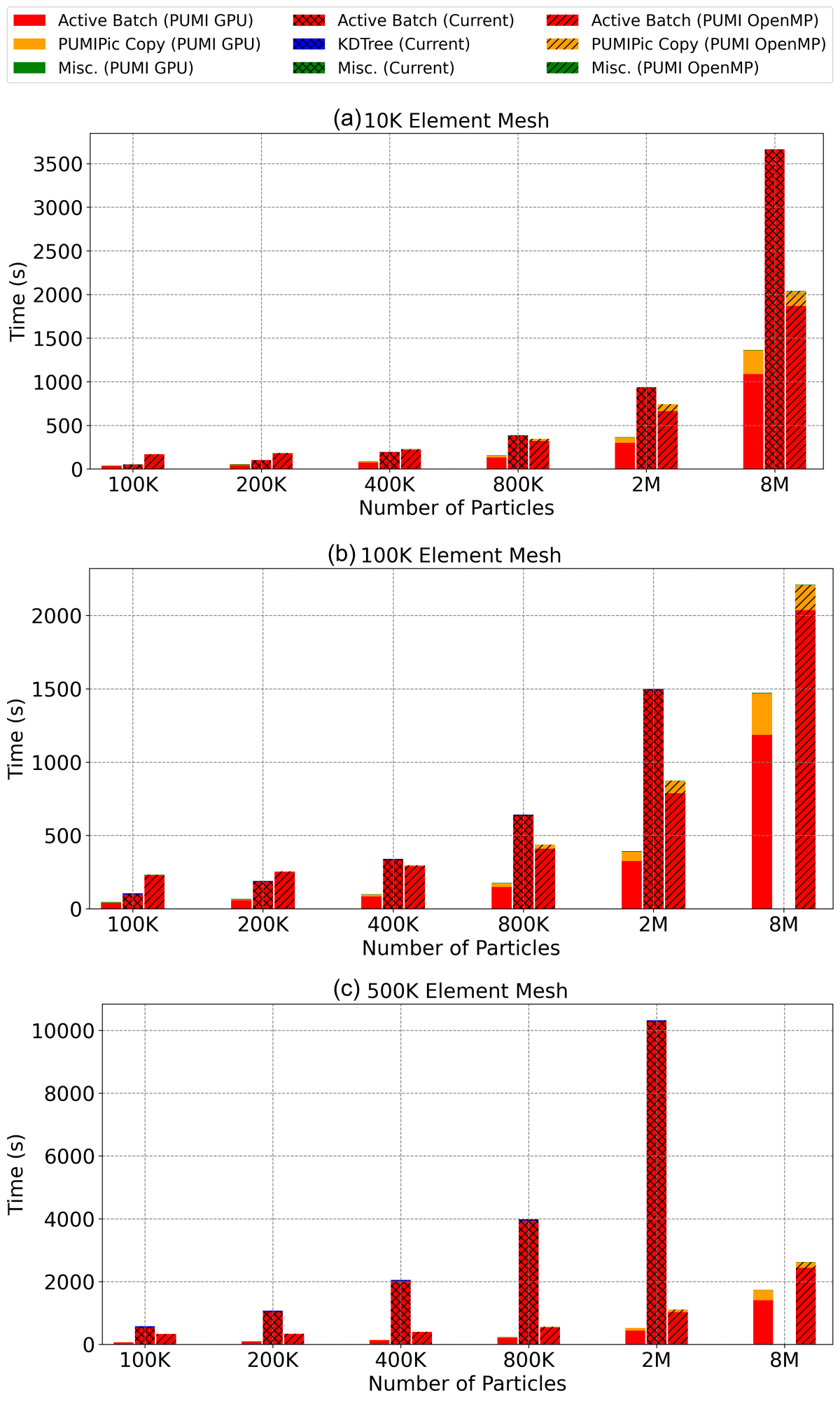}
    \caption{Runtime vs. number of particles for 10,000 elements (a), 100,000 elements (b) 500,000 elements (c). Colors show the proportion of the solution time spent in search initialization (blue), active batch, i.e., transport and tally (red), data copies to the GPU (orange), and other operations (green).}
    \label{fig:scaling-with-particles}
\end{figure}

Since Monte Carlo simulations contain both particles and meshes, it is important to characterize the performance across a range of particle counts and mesh sizes. Figure~\ref{fig:scaling-with-particles} compares the scalability of PUMI-Tally to the current method implemented in OpenMC as the number of particles are increased for three mesh sizes, 10 thousand elements, 100 thousand elements, and 500 thousand elements. The number of particles is increased from 100 thousand to 8 million. The colors in each bar represent the active batch time in red which includes both transport and tally operations. These are timed together because when the tallies are run on the GPU, the tallies can be computed asynchronously. The orange represents the time it takes to copy data from OpenMC to PUMI-Tally, and the blue represents the time it takes to construct the KDTree used for localization in the current OpenMC code.

For the smallest number of particles and elements, the current method is faster than OpenMC with CPU only PUMI-Tally. And, it is about the same speed as OpenMC with GPU based PUMI-Tally. As the number of particles increases, the cost of the current OpenMC solution exponentially increases, whereas PUMI-Tally scales linearly. The difference is exacerbated as the number of particles increases.

Interestingly, the difference between CPU PUMI-Tally and GPU PUMI-Tally is not that large (max of about 2 times). Indicating that much of the performance benefit is coming from the use of batched algorithms and utilization of cache aware algorithms. We believe there is a significant opportunity for additional improvements in the GPU performance through coalescing of active particles to avoid low occupancy when the number of active particles reduces.

We additionally saw that the construction of localization search data structures and copying data from the OpenMC particle data structures to arrays for PUMI-Tally did not take a significant portion of the run time. Results from the current OpenMC method were not collected for the eight million particle cases due to runtime limitations.

\begin{figure}
    \centering
    \includegraphics[width=0.8\linewidth]{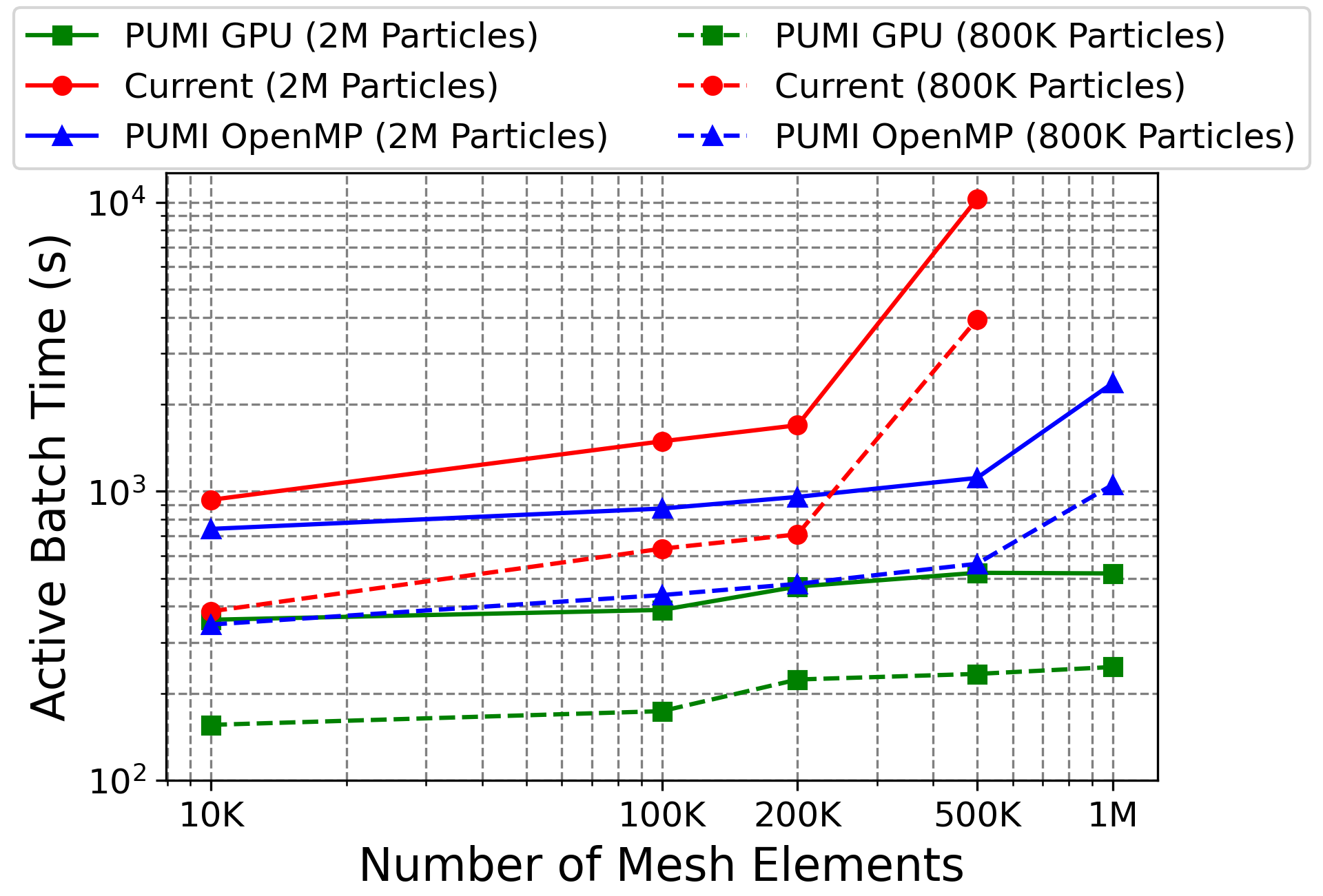}
    \caption{Runtime to simulate 800,000 and 2 million particles vs. the number of elements for the current solution (red circles) and PUMI-Tally (green squares).}
    \label{fig:scaling-with-elements}
\end{figure}

Figure~\ref{fig:scaling-with-elements} demonstrates scaling of the active batch time with the number of elements with 800 thousand particles (dashed lines) and two million particles (solid lines). We observe a similar trend that the current solution scales poorly with the number of elements. The CPU version of PUMI-Tally seems to scale well up to 500 thousand elements and sees a large increase in the cost for one million elements. The GPU version of PUMI-Tally showed relatively flat scaling to one million elements. This suggests that part of the small difference in CPU and GPU performance with PUMI-Tally is related to the GPU having inadequate work.

\begin{figure}
    \centering
    \begin{subfigure}[b]{0.45\linewidth}
        \centering
        \includegraphics[width=\linewidth]{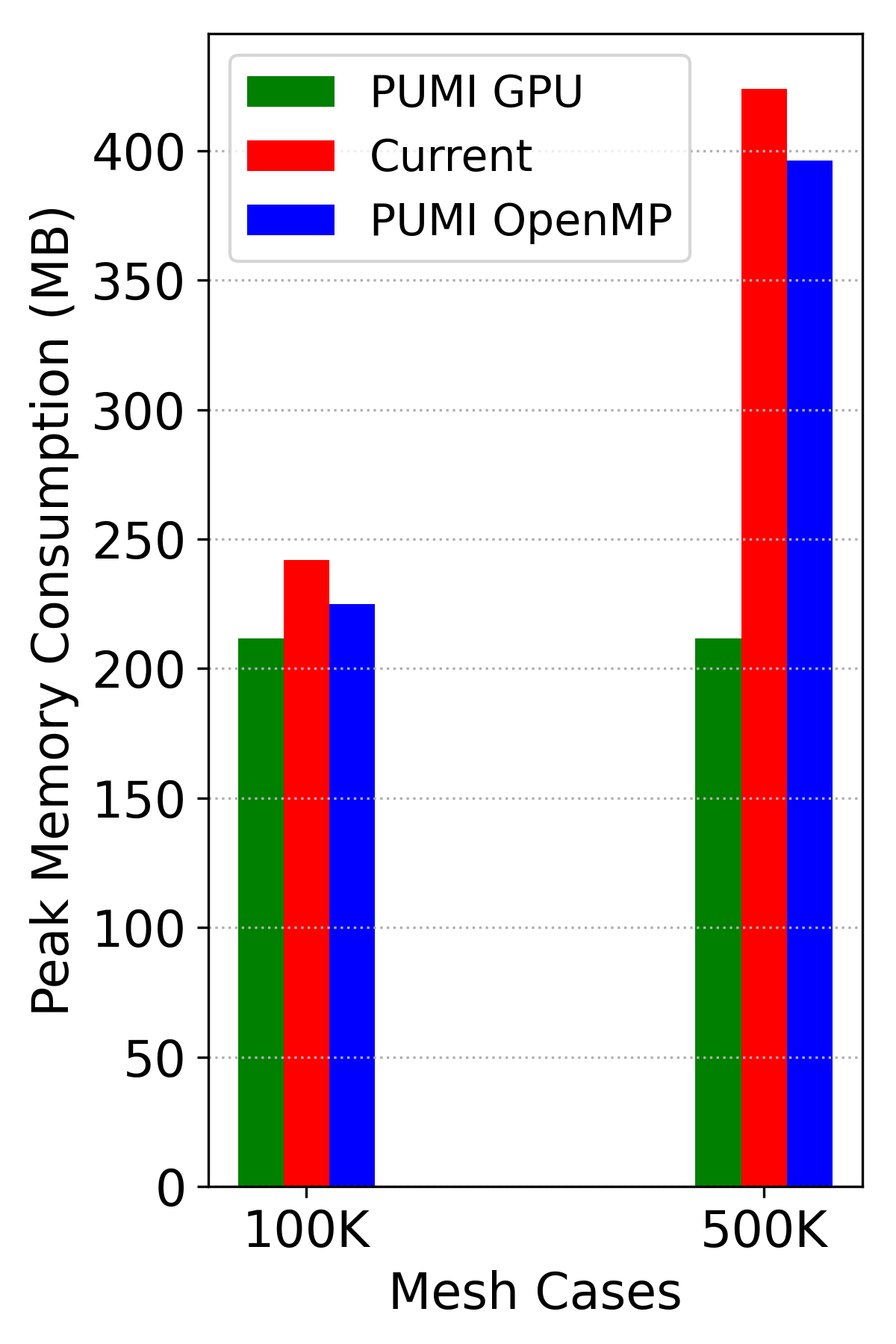}
        \caption{Peak Memory Consumption}
        \label{fig:mem_consumption}
    \end{subfigure}
    \hfill
    \begin{subfigure}[b]{0.45\linewidth}
    \centering
    \includegraphics[width=\linewidth]{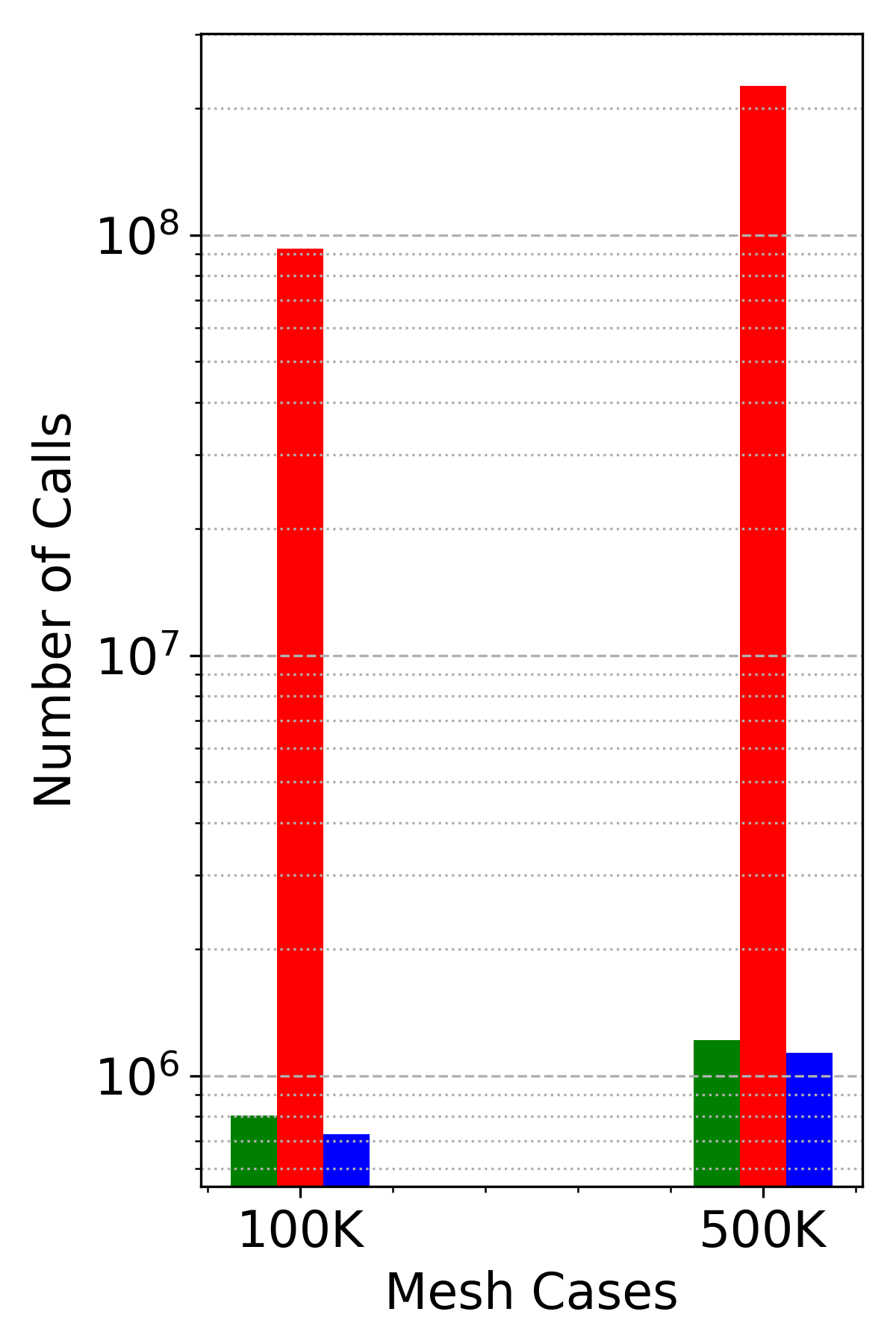}
    \caption{Number of Memory Allocations}
    \label{fig:mem_allocs}
    \end{subfigure}
    \caption{Memory profiling collected on Perlmutter using Heaptrack.}
    \label{fig:memory_profile}
\end{figure}

While testing OpenMC, we noticed that the memory consumption was relatively constant, but was fluctuating. This prompted us to more thoroughly analyze the memory allocations. The memory was profiled on Perlmutter during the simulation initialization and three iterations of the advance particle event. Figure~\ref{fig:mem_consumption} shows the total memory consumed for one hundred and five hundred thousand elements and one hundred thousand particles. Here, we observe that the current method utilizes slightly more memory than the PUMI-Tally solution. The GPU based method utilized less memory because the tallies were not copied to the host during the three iterations over which we performed measurements.

Additionally, we observed that the current OpenMC tallies were performing many small allocations (see figure~\ref{fig:mem_allocs} and note its use of a log scale). And, the number of allocations was increasing with the number of iterations. These allocations come from two places. First, when the particles trace through the mesh, OpenMC records the distance traveled through each element into a vector. Since one cannot know the number of elements a particle will cross a-priori, this requires either a two-pass algorithm, or dynamic allocations. The current solution uses the latter. Secondly, the current design requires that each length that a particle crosses an element is re-localized to the element. This step utilizes the MOAB mesh backend which performs an additional allocation \cite{tautgesMOABSDIntegratedStructured2004}.

The number of allocations that PUMI-Tally makes does not grow substantially with the number of iterations, which is a consequence of the design. In PUMI-Tally, we perform tallying as the particle is transported within the mesh. That is, the particles stop at each element intersection and sum the appropriate tally quantity into the appropriate bin of a multidimensional array based on the element and particle energy. Since the number of elements and number of energy bins are known at initialization time, this approach requires no dynamic allocations. To sum the quantities, we utilize atomics, however there are not typically many particles in the same element with the same energy level, so contention is low.

\begin{figure}
    \centering
    \includegraphics[width=0.9\linewidth]{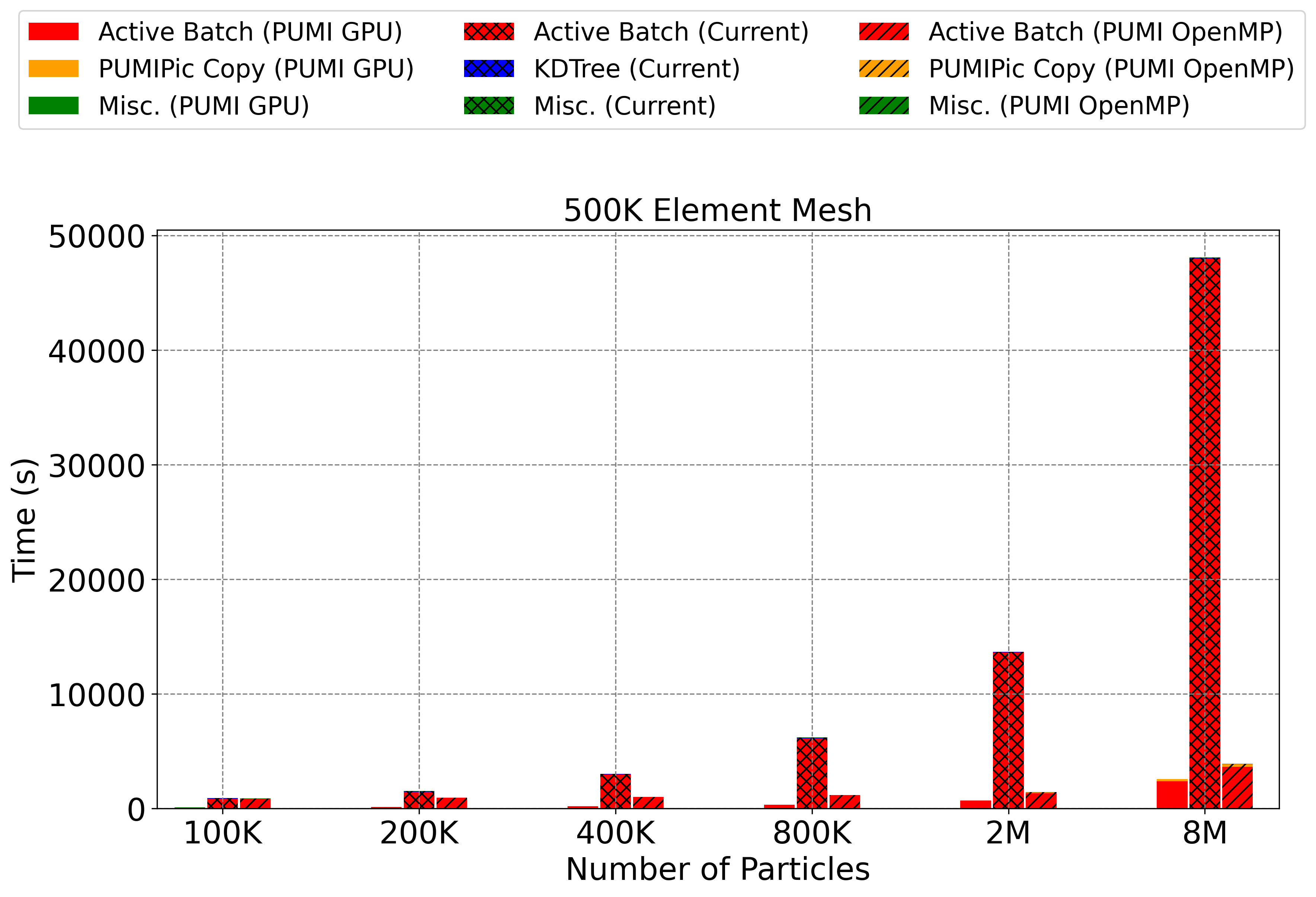}
    \caption{Simulation Time for 500,000 Mesh Element on Empire AI alpha}
    \label{fig:empire-perf}
\end{figure}

Performance results on Empire AI alpha are shown in figure~\ref{fig:empire-perf}. These results are qualitatively similar to those in figure~\ref{fig:scaling-with-particles} showing PUMI-Tally scaling significantly better with the number of particles than the current OpenMC tallies. The overall runtime was longer for both PUMI-Tally and the current method on Empire AI with the current method impacted more than PUMI-Tally. Empire AI alpha is still an experimental system with limited details available to us about the underlying interconnects, networking, etc. We expect that Empire AI has not yet been tuned to the extent that Perlmutter has.

\begin{figure}
    \centering
    \includegraphics[width=0.8\linewidth]{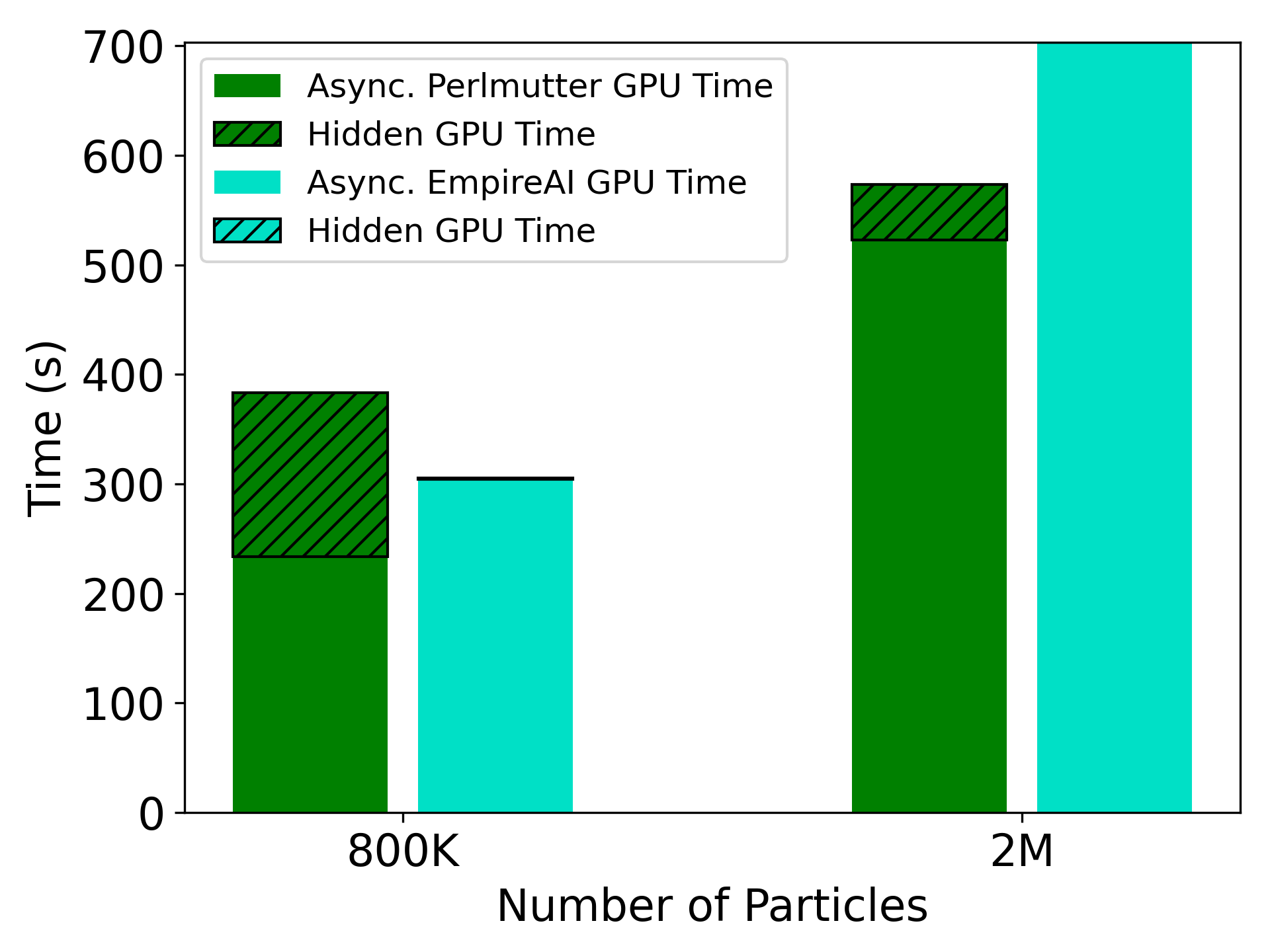}
    \caption{Advantage of Asynchronous GPU Tallying}
    \label{fig:async}
\end{figure}

Since PUMI-Tally runs on the GPU, and OpenMC does not rely on any intermediate tally results, the tallying can be performed asynchronously once the data is copied onto the device. To characterize the impact of this, we ran the tallying with and without a fence around it. In figure~\ref{fig:async}, the total height of the bars is the runtime for the fenced version of PUMI-Tally for a single batch with a 500,000 element mesh. The hashed zone is the difference observed when there is no fence, or the amount of time that is effectively hidden in the overall analysis runtime through asynchronicity. On Perlmutter, a fairly substantial part of the overall runtime was hidden. On Empire AI alpha we observed almost no impact of the asynchronous computation. We believe this is the impact of two factors. First, the H100 is faster than the A100 and second, the time to perform the host to device copy, a synchronous operation, on Empire AI alpha was greater than that on Perlmutter by more than a factor of two.

\begin{figure}
    \centering
    \includegraphics[width=1\linewidth]{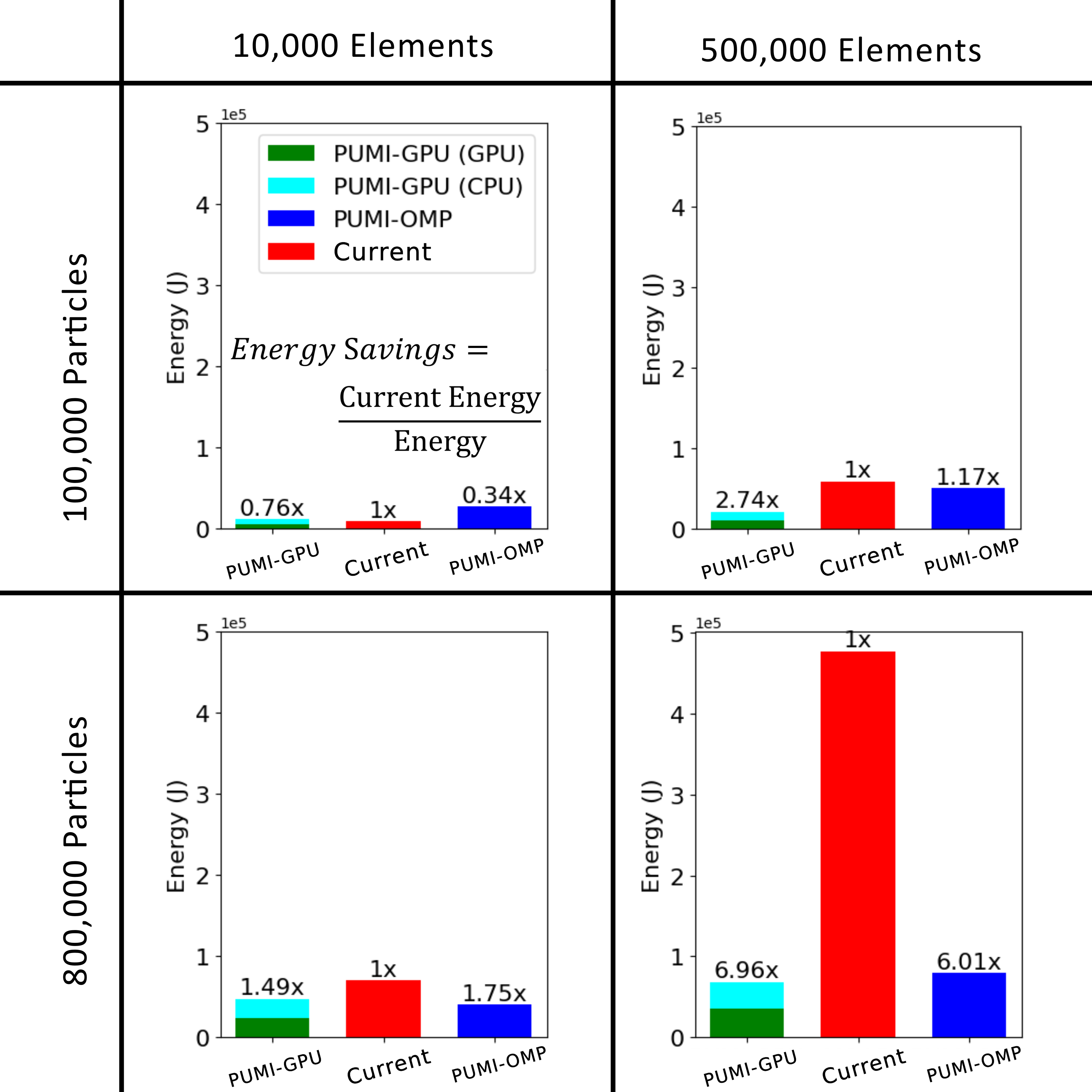}
    \caption{Energy Consumption}
    \label{fig:energy-consumption}
\end{figure}

There has been a recent emphasis on understanding algorithms not just in terms of performance, but also in terms of energy consumption. Figure~\ref{fig:energy-consumption} shows the energy savings with PUMI-Tally over the current OpenMC solution. The energy consumption was measured on Perlmutter using the methods described in \cite{zhaoUnderstandingVASPPower2024} that store power reported by the Cray Power Monitoring Interface into the NERSC Operations Monitoring and Notification Infrastructure (OMNI). The energy savings are defined as \[S = \text{energy current}/\text{energy new}.\] For the GPU version of PUMI-Tally, the bars are split into the portion of energy consumed on the CPUs and GPUs. We observe that for small numbers of particles and elements, the current solution outperforms PUMI-Tally with the CPU version of PUMI-Tally performing worse than the GPU version. As the number of particles and elements increase, PUMI-Tally starts to utilize less energy than the current solution. The combination of increased elements and particles is particularly bad for the current method, with only a minor increase in energy consumption for the other methods. One interesting observation is that the GPU based PUMI-Tally consumes more energy than the CPU version when the number of elements is low, even when the number of particles is high. This indicates that if solving a problem with a low number of elements, one should consider utilizing the CPU solver for a small runtime penalty and a reduction in the energy consumption.

\section{Conclusion}
In this work, we demonstrate the integration of the PUMIPic hybrid particle mesh infrastructure into the open source OpenMC Monte Carlo code to accelerate the computation of tallies on unstructured meshes. This work is critical to support the types of complex models that are needed for fusion simulations. To avoid the need to link OpenMC against a complex set of dependencies that require GPU compilers, we have implemented the PUMI-Tally library which acts to perform Monte Carlo specific operations and split PUMIPic from OpenMC.

The results presented here indicate that our approach is effective when the number of particles or elements is large. Interestingly, we found that our CPU based tallies perform only slightly worse than the GPU based tallies, but substantially better than the current solution. This indicates that the use of batched datastructures, cache aware algorithms, limiting memory allocations, and effective mesh representations are critical even when GPU acceleration is not available. However, when the GPU is available, it can take advantage not only of its computational power, but also the asynchronicity afforded by running an application on both the CPU and the GPU.

We also investigated the power consumption of our algorithms. And, we identified that in some circumstances, when the number of particles and elements are low, it may be worthwhile to utilize alternative algorithms even when they take increased runtime.

Although we demonstrated substantial runtime improvements in the tally operations, more work is needed to support performing the particle transport and tallies simultaneously. This will allow executing the full analysis workflow on the GPU accelerator as well as optimization of the layout of active particles when there are many inactive particles towards the end of the simulation.

\begin{acks}
\Redacted{
This research was supported by the U.S. Department of Energy, Office of
Science, under awards DE-SC0021285 (FASTMath SciDAC Institute) and
DE-AC02-09CH11466 (StellFoundry: High-fidelity Digital Models for
Fusion Pilot Plant Design). This work used the resources of the National Energy Scientific Computing Center (NERSC) at the Lawrence Berkeley National Laboratory. Any opinions, findings, and conclusions or
recommendations expressed in this material are those of the author(s)
and do not necessarily reflect the views of the U.S. Department of Energy. We gratefully acknowledge use of the research computing resources of the Empire AI Consortium, Inc, with support from the State of New York, the Simons Foundation, and the Secunda Family Foundation.}
\end{acks}

\printbibliography


\end{document}